\documentclass[
aps,%
12pt,%
final,%
notitlepage,%
oneside,%
onecolumn,%
nobibnotes,%
nofootinbib,%
superscriptaddress,%
noshowpacs]%
{revtex4}
\usepackage{graphicx}
\usepackage{color}

\begin{document}

\noindent {\it Astronomy Reports, 2026, Vol. , No. }
\bigskip\bigskip  \hrule\smallskip\hrule
\vspace{35mm}


\title{The Crab Nebula progenitor: recovering the 1054 AD supernova event as galactic Gamma-ray burst.}

\author{\bf \copyright $\:$  2026.
\quad \firstname{C.}~\surname{Sigismondi,}}%
\email{sigismondi@icra.it}
\affiliation{ICRANet, Pescara, Italy}%

\author{\bf \firstname{R.}~\surname{Ruffini.}}
\affiliation{ICRANet, Pescara, Italy}

\begin{abstract}

\centerline{\footnotesize Received: ;$\;$
Revised: ;$\;$ Accepted: .}\bigskip\bigskip\bigskip

In 1054 AD a daytime star appeared in the constellation of Taurus, for three weeks, and it was reported in various sources from Europe to China/Japan: it was one of the few documented galactic supernovae of the last two millenia.
This paradigm has been established about sixty years ago, as the comprehension of the physics of supernovae progressed with enough observational data. The Gamma-ray bursts were discovered in the same period, but only in the past few years have their observations become  daily and their distances have been fully understood as cosmological. After the explosion, the exponential decay of the luminosity in gamma-rays and X-rays has been followed with telescopes onboard dedicated satellites. Also the exponential decay of the afterglow's optical and radio frequencies have been observed with the largest optical and radio telescopes.
Within the binary-driven hypernova framework, successful in explaining all the observed phases of the Gamma-ray bursts, the universal exponential decay can be extended to 1000 years after the burst, to account for the present values of Gamma and X-rays as well as optical and radio frequencies of the Crab Nebula.
Both the daytime visibility of the burst, and the simultaneous radiation plagues appeared in Constantinople and Cairo is a strong evidence of the presence of Gamma-rays in the lower atmosphere, coming from the same source originating the Crab nebula. The association to the daytime visibility of that star and the following plague meets exactly the etymology of the word dis-aster, bad star. 

\end{abstract}

\maketitle

\section{Introduction}

The rate of supernovae per century of the Milky Way is statistically larger than the four ones (1006 in Lupus, 1054 in Taurus, 1572 in Cassiopeia and 1604 AD in Ophiucus)\cite{1} observed in the last millenium.  Nevertheless the galactic dust on the galactic plane may have hidden some of them (1680 AD in Cassiopeia).\cite{2}
The number of Gamma-ray bursts (GRBs) observed, the most luminous sources of the Universe, is now around one per day, with the most luminous appearing from cosmological distances $z\sim4$.\cite{3} 
The enormous energy released by such an event is associated with the beamed emission of Gamma-rays in quantities sufficient to penetrate temporarily our atmosphere, destroying life, if the distance is below 15 Kpc.\cite{4}
The Crab nebula, which was produced by the 1054 AD event, is at 2 Kpc from Earth.
If the Crab nebula was originated by a GRB,\cite{5} the beamed radiation should have been slightly deviated from Earth, so that the "disaster" did not cancel completely the life in the burst-side Earth's hemisphere. 
The word "disaster" means literally a "bad star", which actually was observed also in daytime, and recorded by the chroniqueurs. 

\section{Historical grounds}

The study of the historical sources has been extensive, and chronological inconsistencies have been evidenced: from Chinese (4 July) to Japanese (August) sources there are several weeks of difference in the time of appearance. The European chronicles offer a different chronology: Oldemburg, Rome 19 April, Constantinople and Cairo (June-July).
Moreover, the original source of the Chinese chronicles, upon which the event is known as "Chinese supernova", has been transcripted more than two centuries after the event, when a new dinasty, Yuan, took power. The astronomical observations were strictly linked to the interpretation of celestial signs referred to the king, and a bad omen (event, celestial location and time) toward a king or the whole dinasty should have been avoided from an exact report. 
In general, all the ancient chronicles may have suffered some censorship or interpreting bias, which influenced the report of a precise date.
We already presented a detailed analysis of the sources\cite{5} and the hypothesis of a conventional 3-July chronology, since the explosion of the 1054 supernova is traditionally considered on 4 July, accepting one the Chinese chronicles as chronologically correct. 
The explosion occurred with the Crab rising for the Americas and around the meridian transit for Europe. In Europe, with the lower airmass, it produces injuries and casualties for radiations, among the people exposed to them (Constantinople and Cairo). Few hours later the Crab-source rised also in China where it was observed from dawn to daytime for 21 days, and followed until it disappeared for two years.
The people in China were in the hemisphere opposite to the burst, at the moment of the explosion, without suffering radiation damages, since the exponential decay lowered greatly the Gamma-rays luminosity of the source.
The Chinese astronomers observed the optical afterglow's decay, allowing the present astronomers to identify accurately the position of the source as the one which originated the Crab nebula.

\subsection{The Sphere of Light of Oudenburg}

\begin{figure}
\includegraphics[width=1\textwidth]{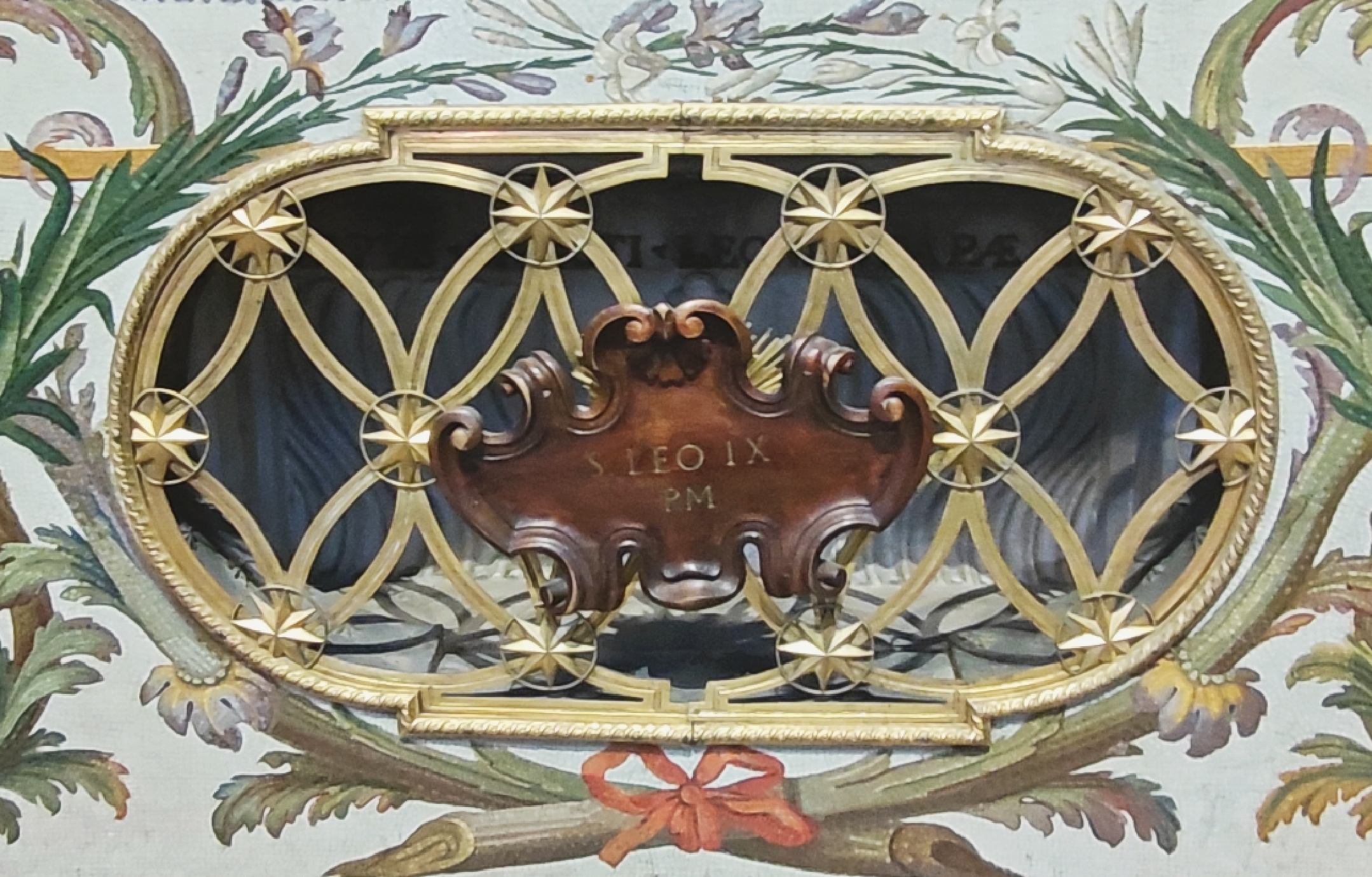}
\caption{Corpus Sanctis Leonis Papae IX. The body of S. Leo IX pope, death on 19 April 1054 AD, at the time of the Crab nebula originating event. His body is in the ancient scraped sarchophagus (roman, reused), embodied by the new altar realized in the style of the others, consecrated by pope Benedict XIII in 1729.}
\label{bin1}
\end{figure}

\begin{figure}
\includegraphics[width=1\textwidth]{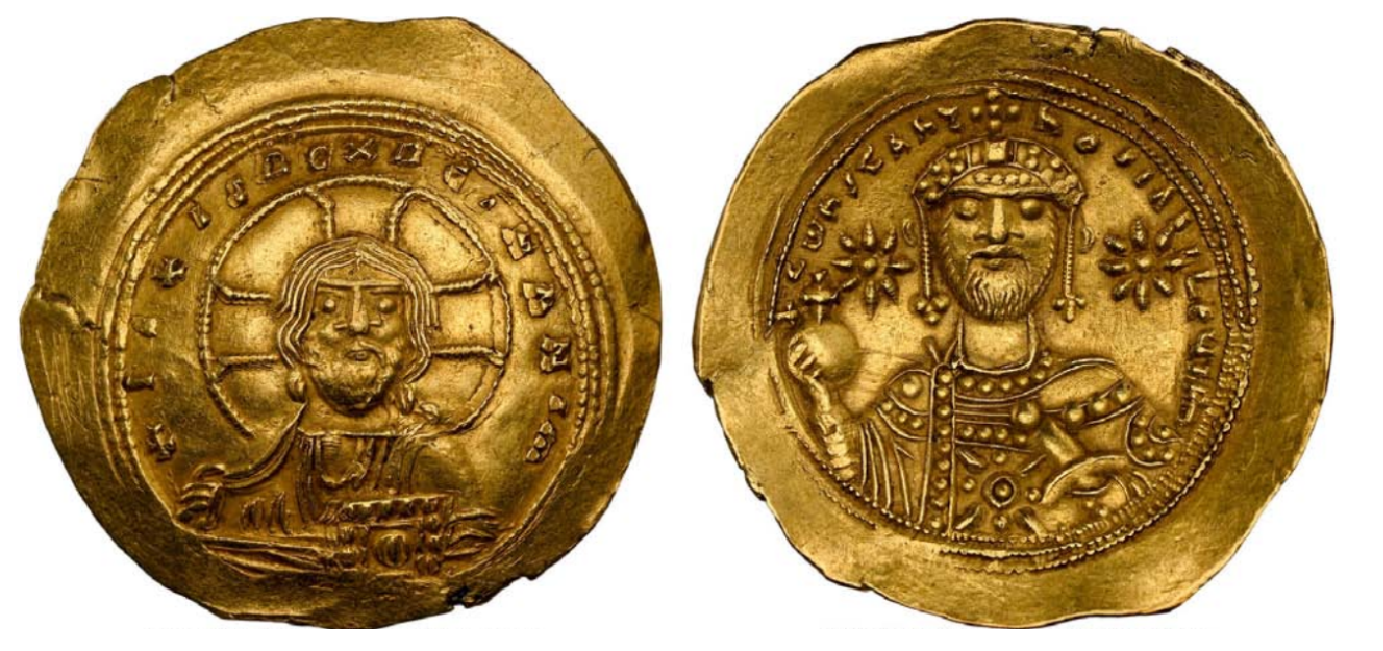}
\caption{A rare coin issued by the Emperor Constantine IX Monomachus (11 June 1042 June- 11 January 1055 AD) where it could have been represented the Crab nebula originating event, with a star repeated two times in the field.}
\label{bin2}
\end{figure}

The escription of a sphere of light seen in Oudenburg appeared in the chronicle of St. Peter's abbey, and it has been already presented as a demonstration that the source was pointlike.\cite{6}
Here we add a reconstruction of the historical circumstances described in that chronicle.
The observation: a sphere of light was observed near noon, with the Sun high in the sky.
The interpretation and the (consequent) chronology: it was the soul of the saint pope Leo IX (dead on 19 April 1054) rising to heaven.
The implicit date of this chronicle is then 19 April, since the soul of a saint is supposed to rise to heaven immediately after the death.
Leo IX was German and travelled also in the area of Oudenburg (Flandres, modern Belgium), so he reliably was known personally by that community.
Pope Leo IX was buried in the Paleochristian Basilica of St. Peter in the Vatican, and now, after the Renaissance's renovation, his body is in the St. Joseph transept, in the right altar, (see Fig.~ \ref{bin1}).
The importance of Leo IX in the storiography of that astronomical event is twofold.
1. the observation of the fireball in daytime, and 2. the separation occurred after his death of the Eastern (Orthodox) Church, from the Western (Catholic). Leo IX gave the faculty to his legates to excommunicate the patriarch of Constantinople Michael Cerularius, and this happened on 16 July, nearly three months after the death of Leo IX.
The emperor of Constantinople was Constantinus IX Monomachus, and among his yearly coins there is only one with a star, which has been considered reproducing the daytime star visible during that period
(see Fig.~ \ref{bin2}). Moreover, Constantinus IX died at the beginning of the following year, 1055 AD.
Numismatic studies tend to exclude now the astronomical interpretation of the coin's star, since the presence of a star is a rather common fact. 
Nevertheless, the coincidence of the astronomical, political, religious and epidemiological events is still remarkable.\cite{5}

\subsection{The Plagues in Egypt and Constantinople}

Under the generical name of plagues there were various type of diseases, bacterial or viral. If the origin would have been radiational the name used at that time would have been the same. A great number of people died after the apparition of that star for a plague in Cairo and in Constantinople.
The hygienic conditions may have spread also other plagues, but the connection with the star is explicit in the accounts of the physician Ibn Batlan.\cite{2}
Our interpretation is concerning the radiational nature of the plague, like after the A-bomb in Hiroshima and Nagasaki many people suffered the consequences to the exposition to the bomb direct (Gamma-rays) radiation. This disease lead people to death in a few days.
Therefore, the plagues after the apparition of the new star are connected with its abundant Gamma-rays radiation penetrated into the atmosphere down to the ground level.
The atmosphere is naturally blocking X-rays and Gamma-rays, but when the radiation flux is overwhelming its barrier is overcome and the radiation arrives to the sea level.
A proof of this fact may be encountered in the ice cores of South Pole, where the presence of the isotope $O^{18}$ trapped in the former atmosphere and then in the ice core, has been measured binning 20 years per epoch. The Oort minimum of the Sun as the other grand minima, coincided with a rising of cosmic rays received down to Earth, so the pointlike signal of the Crab event is in phase with this enhancement, and -because of the 20-years binning- it is comparable with the bidecadal dose of cosmic rays at solar minimum. 

\subsection{The Visibility in Daytime}

The third fact supporting the GRB (but also the classical supernova) hypothesis is the visibility of the fireball in daytime, followed by the  21-days visibility of the new star in daytime, and up to two years (630 days) to the naked eye in nighttime.
The optical afterglow of GRB decreases exponentially from initial values ranging around $10^{54}$ erg/s, for the most luminous ones. 
Also, the novae and the supernovae undergo an exponential decay, right after the photospheric phase of the explosion; their spectra loose the thermal continuum, acquiring the nebular features as the ejected material expands into the surrounding interstellar medium through a shock wave with decreasing density. The decay for the novae is as fast as 3 magnitudes in 10 days, afterwards the general slope decreases.

\begin{figure}
\includegraphics[width=1\textwidth]{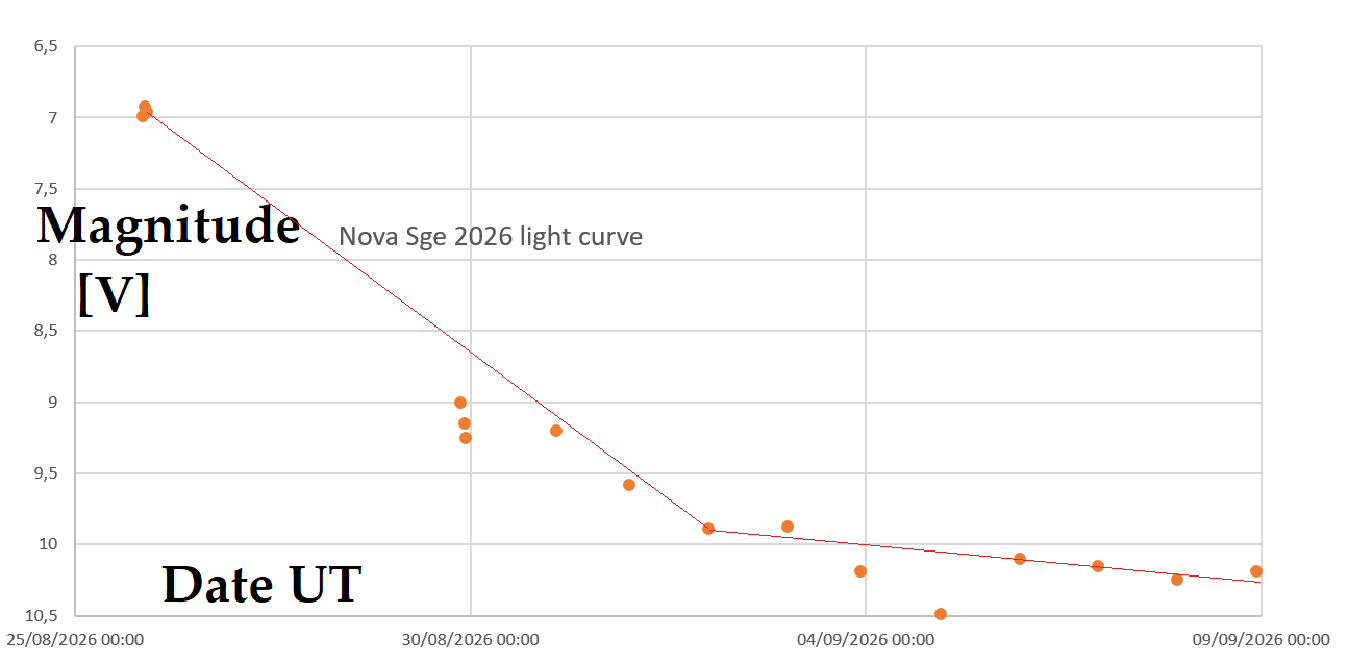}
\caption{The luminosity of the Nova Sagittae 2026 from AAVSO catalogue with the photometric measures of C. Sigismondi (SGQ), obtained at the Asiago Astrophyscal Observatory of the University of Padova 122 cm Galileo telescope, and with the PHYSIS refracting telescope 90/500. The exponential decay shows two slopes. The same situation may have happened in 1054 AD during the first weeks of daytime visibility of the Crab nebula progenitor.}
\label{bin3}
\end{figure}

In Fig.~3 \ref{bin3} the initial decay of the Nova Sagittae 2026 (exploded on 25.12 UT August 2026) is plotted.

This is typically the occurrence of a nuclear ignition on the surface of a white dwarf, which may somewhat reproduce, at much lower energy scales, the GRB supernova, as the binary-driven hypernova model suggests.\cite{3}
In order to be easily visible in daytime a star has to shine below magnitude -5, then being brighter than Venus.
A drop of 3 magnitudes in 21 days, like a classical nova (not fast) may show, has to reach magnitude -5, then the initial value could have been -8 mag in optical light, for the Chinese observers.
This luminosity is still not enough bright to produce a daytime halo, being the observed fireball an atmospheric halo with precise angular dimensions of 22 degrees of radius, the angle of distance of a parhelion from the Sun, or the radius of a full solar halo.
It is well known that the Moon produces haloes during the night under particular conditions, but this halo would not be enough bright to be visible in daytime.
In order to see an halo or a sphere, as it is described in the Oudenburg's chronicle, the luminosity of the source should be comparable with the Sun, i.e. mag. -26.4.
Then the star in daytime has been the afterglow decay, while the sphere of light in daytime has been the burst itself, nearly 20 magnitudes brighter, or $10^8$ times more luminous.
In this phase the Gamma-ray component of the spectrum penetrated in the atmosphere was mortal or extremely dangerous for the skin, in absence of clouds and veils.
In Oudenburg the visibility of the halo was due to high clouds, which prevented the Gamma-rays to penetrate down to the ground and then exempted the witness of the phenomenon from a rapid onset of a radiation skin disease. 

\section{Timing and Energetics of the phenomenon}

The historical documents are controversial about a precise timing for several reasons.
As in the previous paper\cite{6} we unify the chronology of the Crab progenitor event.
Explosion seen simultaneously in the Americas at its rise and in Europe in the middle of the sky.
In Oudenburg the halo was visible. The luminosity was comparable to the Sun. In the first hour the Gamma-rays radiation was deadly, if no veils filtered it. 
The Crab nebula is located at 2 Kpc from us, in the Milky Way.
At 2 Kpc of distance a flux comparable with the solar one received at Earth $Phi_{odot}=1360 W/m^2$ corresponds to a luminosity of $6.5 \times 10^{49} J/s$ or $6.5 \times 10^{56} erg/s$ which is the range of a peta-nova ($10^{54} erg/s$). 
The sphere of light accounted in the Oudenburg chronicle was visible in daytime without exceeding the solar luminosity, then the peta-nova range is plausible.
Moreover the beaming of the GRBs makes it possible that not all the energy flux released by the GRB invested the Earth's atmosphere provoking victims only where the sky was clear and the Sun at meridian (Cairo and Constatinople). 
Since all the aforementioned reasons the chronology of the explosion in 1054 AD as 4th July is merely conventional, and since the occurrence of cloud veils and haloes is more frequent in Spring than in Summer time, a chronology starting from 19 April 1054 AD is also convincing.  
Considering the Gregorian reformation of the Calendar, 19 April 1054 AD corresponds to 26 April of our calendar, with the Sun at only 36 degrees from the Crab nebula, exactly the same angular distance of the 4th July 1054 AD, when the 25-days waning Moon was close to it and to Aldebaran. This last detail, the Moon, is present only in the Arizona drawings, while it is mentioned only as a position's matter (first day new Moon's position of the new star) in another chronicle of Bologna.\cite{5}
The association of the Moon with a star in the flags of Muslims is linked to the siege of Constaninople which occurred later in 1453, and probably has nothing to do with the apparition of the Crab nebula progenitor.

\section{Conclusions}

With respect to our previous publication\cite{5} which introduced the idea that the Crab nebula was originated by a GRB and gave a unified chronology of the various accounts from Arizona Indian drawings to Oudenburg, Rome, Cairo and Constantinople in addition to Chinese and Japanese, and to the other publication\cite{6} that treated the Oudenburg chronicle as the account of a GRB simulating also the atmospheric halo, we here added four new points:

1. To have a daytime halo visible, we need a source's luminosity comparable with the Sun, and at 2Kpc distance this correspond to a peta-nova.
2. The beaming factor of the GRB and the presence of cloud veils may have limited the radiation injuries to the population exposed to it, reliably in the late Spring/Summer months of 1054 AD. 
3. The pope s. Leo IX's burial place, whose death was associated to the observation of the atmospheric halo in daytime different from the solar one and to the forthcoming abrupt separation between Orthodox and Catholic, a true dis-aster for the Church.
4. The optical decay of a fast nova (Nova Sagittae 2026) showed a slope of 3 magnitudes in 10 days. Applying such decay to the optical afterglow, visible in daytime, we obtain a starting point at magnitude around -8, which is not enough luminous to produce a daytime halo (it is luminous as a Moon crescent).  

In the previous paper\cite{5} we evidenced also that the energetics of the GRB decay in the channels of X-rays, Gamma-rays and Radio, may lead -extrapolated- to the present values of the Crab radiations.

The application of the binary-driven hypernova models to fit successfully the GRB 220101A and the other peta-novae events\cite{3} predicts the production of a pulsar and of a black hole. The observation of the Crab pulsar made in 1967 is in agreement with that model. The black hole immersed in the remnant, still to be found.

The key point to validate a galactic GRB, exploded within our safety zone,\cite{4} as the Crab nebula progenitor is the atmospheric halo observed in Oudenburg, which was associated -in the chronicle- to the assumption of the soul of the defunct pope Leo IX, and the casualties for radiation damages described as plagues by the ancient physician Ibn Batlan.




\clearpage


\clearpage


{\bf Figure captions to Sigismondi and Ruffini}
\bigskip\bigskip

Fig.~1.~Corpus Sanctis Leonis Papae IX. Body of S. Leo IX pope, death on 19 April 1054, at the time of the Crab nebula originating event. His body is in the ancient scraped sarchophagus (roman, reused), embodied by the new altar consecrated by pope Benedict XIII in 1732. 

\bigskip

Fig.~2.~The yearly coin of the Emperor Constantine IX Monomachus, contemporary of the Crab nebula originating event, with a star in the field.

\bigskip

Fig.~3.~The luminosity of the Nova Sagittae 2026 from AAVSO catalogue with the photometric measures of the author (SGQ).

\clearpage


\end{document}